\documentclass[usenatbib]{aa}

\usepackage{graphicx}
\usepackage{amssymb}
\usepackage{epsfig}
\usepackage{psfig}
\usepackage{natbib}
\def\comptt{{\sc comptt}}

\def\nh{{$N_{\rm H}$}}
\def\dbb{{\sc diskbb}}

\def\gx{4U~1728-34}
\def\be{\begin{equation}}
\def\ee{\end{equation}}

\begin{document}

\title{The X-ray spectrum of the bursting atoll source 4U~1728-34 observed with
INTEGRAL}
\author{M. Falanga\inst{1,2}\fnmsep\thanks{\email{mfalanga@cea.fr}},
D. G\"otz\inst{1},
P. Goldoni\inst{1,2},
R. Farinelli\inst{3},
A. Goldwurm\inst{1,2},
S. Mereghetti\inst{4},
A.~Bazzano\inst{5},
L.~Stella\inst{6}
}
\offprints{M. Falanga}
\titlerunning{INTEGRAL observation of 4U~1728-34}
\authorrunning{M. Falanga et al. }

\institute{CEA Saclay, DSM/DAPNIA/Service d'Astrophysique (CNRS FRE
  2591), F-91191, Gif sur Yvette, France
\and Unit\'e mixte de recherche Astroparticule et
Cosmologie, 11 place Berthelot, 75005 Paris, France
\and Dipartimento di Fisica, Universit\`a di Ferrara, via Paradiso
  12, 44100 Ferrara, Italy
\and INAF--Istituto di Astrofisica Spaziale e Fisica Cosmica, Milano,
via Bassini 15, 20133 Milano, Italy  
\and INAF--Istituto di Astrofisica Spaziale e Fisica cosmica, via del
Fosso del Cavaliere 100, 00133 Roma, Italy 
\and INAF--Osservatorio Astronomico di Roma, via Frascati 33, 00040 
  Monteporzio Catone (Roma), Italy 
}

\abstract{We present for the first time a study of the 3--200 keV
  broad band spectra of the bursting atoll source \gx\ (GX~354-0)
along its hardness intensity diagram. The analysis was done using the
{\em INTEGRAL} public and Galactic Center deep exposure data ranging from
February 2003 to October 2004. The spectra are well described by a
thermal Comptonization model with an electron temperature from 35 keV  to 3
keV and Thomson optical depth, $\tau_{\rm T}$, from 0.5 to 5 in a slab
geometry. The source undergoes a transition from an intermediate/hard
to a soft state where the source luminosity increases from 2 to 12\%
of Eddington. We have also detected  36 type I X-ray bursts two of
which show photospheric radius expansion. The energetic bursts with
photospheric radius expansion occurred at an inferred low mass
accretion rate per unit area of $\dot m \sim1.7\times10^{3}$ g
cm$^{-2}$ s$^{-1}$, while the others at a higher one between
$2.4\times10^{3}-9.4\times10^{3}$ g cm$^{-2}$ s$^{-1}$. For \gx\ the
bursts' total fluence, and the bursts' peak flux are anti-correlated
with the mass accretion rate. 
The type I X-ray bursts involve pure helium burning either during the hard
  state, or during the soft state of the source.

}
\maketitle

\section{Introduction}
\label{sec:intro}

Low-mass X-ray binary  systems (LMXBs), consisting of an
accreting neutron star (NS) and a main-sequence donor star (M$<$
1 M$_{\odot}$), are usually classified as atoll or Z sources, according
to the path they describe in an X-ray colour-colour diagram (CCD) or
hardness-intensity diagram (HID). The main difference between atoll
and Z sources is in the accretion rate, the magnetic field and
probably in the spin rate and the geometry of the system.  For
atoll sources, the 
positions along the characteristic branches have historically been
called the ``island'', ``lower banana'' and ``upper banana'' branches, which 
indicate the spectral states probably  as a function of the 
mass accretion rate \citep{hasinger89}. The mass accretion rate  increases
from the island state (hard spectral states) where the soft emission
is much reduced and the spectrum is dominated at high energies up to
200 keV, to the banana states (soft spectral states) where most of the
energy is emitted below $\sim 20$ keV \citep[e.g.,][]{barret01,gd02}.

The atoll sources show luminosities smaller than those of Z sources
(generally less than 10 
per cent of Eddington, $L_{\rm Edd}$), low magnetic fields
(B$\sim10^{8}$G), and spectral variability,  or equivalently, motion along the
banana branch over a timescale of hours to days, and more slowly in the
island state from days to weeks.   

Some atoll sources also show  time variability properties in their
power spectra, such as Quasi-Periodic Oscillations (QPOs), which are
strongly correlated with the position on the CCD/HID \citep[][]{hasinger89}. 
Almost all the timing features characterizing the 
power density spectra of these sources seem to vary in a smooth and
monotonic way when the source moves along its CCD/HID
\citep{wvdk99,pbvdk99},  
implying a tight correlation between  spectral and temporal  behaviour
\citep[for a review see][]{vdk04}. In most atoll sources, QPOs are
observed in the frequency range from  millihertz to the kilohertz. 
Kilohertz QPOs or twin kHz QPOs in the frequency range from
300--1200 Hz are usually observed in the island state and lower banana
branch \citep[][and references therein]{vdk04}.      

In some of the atoll sources, type I X-ray bursts
are also observed \citep[e.g.,][]{lewin83}. Type I X-ray bursts are
characterized by a fast rise and exponential 
decay with durations ranging from seconds to tens of minutes. These
bursts are due to unstable hydrogen/helium burning in a thin shell on
the NS surface \citep[see e.g., the review by][]{lewin93}. The burst
spectrum can be described by blackbody radiation with cooling during
the decay of the burst. The gradual decay persists longer at lower
energies, indicating that the burst spectrum is
characterized by steadily 
decreasing blackbody temperatures \citep[e.g.,][]{lewin83}. 
During these bursts nearly-coherent oscillations are sometimes observed,
the frequencies of which are in the rather narrow range between 300 and 600~Hz 
\citep[][]{sm99,s04}. This frequency is 
interpreted as the NS rotation frequency  due 
to a hot spot (or spots) in an atmospheric layer of the rotating NS
\citep{cmm03}.

The CCDs diagrams and HIDs are powerful ways of parameterizing
the spectral changes using physically motivated spectral models to
understand the underlying physical changes in the source emission.  
In this work, we study for the first time the broad band
(3-200 keV) spectral shape of the atoll source \gx\ as a function of its
HID position. The correlation of type I X-ray  burst properties
with source state in the HID is also investigated. We analyze the
{\it International Gamma-Ray Astrophysics Laboratory} ({\em INTEGRAL})
2003/2005 data concentrating  also on the high energy 
emission of \gx. This paper deals only with type I X-ray bursts,
therefore whenever we write burst(s), we mean type I X-ray burst(s).   

\subsection{The source 4U~1728-34}

The persistent LMXB source \gx\ (or GX~354-0) is a well
known burster 
discovered by Uhuru and classified as an atoll source from its colour-colour 
diagram and  timing properties \citep{forman76,hoffman76,hasinger89}. 
However, little is known about the system as the optical counterpart
is not identified due to the high optical extinction in the direction
of the  Galactic center. The distance is poorly known, the current estimate
being between 4.4--5.1 kpc using photospheric radius expansion burst
luminosity as a standard candle \citep{salvo00,galloway03}. 
This is based on the assumption that the bolometric burst peak luminosity
during photospheric radius expansion is saturated at the Eddington limit. 
  
The time averaged persistent soft ($0.1$ keV) to hard (300 keV) X-ray
emission spectrum of \gx\ was studied in the past using {\it Einstein},
{\it SAS 3}, {\it EXOSAT} {\it SIGMA} and {\it ROSAT} data
\citep{gh81,basinska84,w86,claret94,schulz99} and, more recently, 
using {\it RXTE}, {\it BeppoSAX}, {\it ASCA} and {\it Chandra}  
\citep{salvo00,piraino00,narita01,dai06}.
The spectrum  was well fitted by a thermal
bremsstrahlung or thermal Comptonization model as well as a Gaussian
emission line at the energy $\sim$ 6.7 keV. Different spectral
parameters were found for the electron temperature and Thomson optical
depth, which can be explained by the varying state of
the source in its CCD/HID position.  

During all spectral states, bursts of $\sim$ 10--20 s duration were
observed \citep[e.g.,][]{cornelisse03}. The burst spectrum can be fitted by a
blackbody with temperature increasing from 1 to 3 keV during the burst
rise time. Most of these bursts show evidence for photospheric radius
expansion
\citep{hoffman76,w86,day90,foster86,franco01,vss01,galloway03,galloway06}.    
Flux oscillations at 363 Hz during the bursts were  discovered in this
source with the {\it Rossi X-ray Timing
  Explorer} ({\it RXTE}), and interpreted as the spin frequency of the
NS \citep{strohmayer96,cmm03}. Apart from the bursting activity, \gx\
also exhibits complex behaviour on short time scale. It was also the
first neutron star LMXB to show twin kHz QPOs in the persistent
emission with a frequency separation almost constant at $\sim 350$ Hz
\citep[e.g.,][and references therein]{mendez99,salvo01,migliari03}.

\section{Observations and Data}
\label{sec:integral}

The present data set was obtained with the {\it INTEGRAL} observatory
\citep{w03} using the 2003/2005 Core Program Galactic Center Deep
Exposure observations and publicly available data. 
%The publicly available observations were performed in the following
%periods: Feb. 28--Mar. 2, Apr. 6--7, Apr. 14--15, Apr. 21--22,
%Aug. 23--29 and from Aug. 30--Sep. 24, 2003. 
We analyzed data from the  Imager on Board the Integral Satellite (IBIS)
 coded mask INTEGRAL Soft Gamma-Ray Imager (ISGRI)
 \citep{u03,lebr03} at energies between 20 and 200 keV and
from the JEM-X monitor, module 1 and 2 \citep{lund03} between 3
and 20 keV. For ISGRI and JEM-X, the data were extracted for all
pointings within  $\leq$ $3\fdg5$ from the source direction. The available data
consist of 356 individual pointings for a total effective exposure of 640 ks.
Data reduction was performed using the standard Offline Science 
Analysis (OSA) software version 5.0 distributed by the {\it INTEGRAL}
Science Data Centre (ISDC) \citep{c03}. The algorithms used in the IBIS/ISGRI
analysis are described in \citet{gold03}.

The brightest bursts were detected using the {\it INTEGRAL} Burst
Alert System (IBAS) software \citep{mereghetti03}  running at ISDC.
IBAS is dedicated to the real time discovery and localization of
gamma-ray bursts, transient X-ray sources and bursts in the IBIS/ISGRI
data stream. Bursts are identified by scanning the data binned using
different timescales (from a few milliseconds to 100 s) and
energy ranges. One of IBAS operation modes, running on the
15--40 keV energy interval and 10 s time scale, is particularly
suited to detect type I X-ray bursts. 
Typically the bursts are localized with a $\sim3'$ uncertainty. 
The fainter bursts were detected using 2 s rebinned JEM-X light
curves in the 3--20 keV energy band. Most of the bursts detected with
JEM-X were also detected with ISGRI in the 18--40 keV energy band,
these flux excesses were confirmed also by the imaging as related to 
the source position. So, we call the bursts detected with IBAS
``bright bursts'' and the ones detected only with JEM-X ``weak bursts''.

The burst light curves are based on events 
selected according to the detector illumination pattern for \gx; for 
ISGRI we used an illumination threshold 
of 0.6 for the energy range 18--40 keV.  For JEM-X we used the event
list of the whole detector in the 3--6 keV, 6--12 keV and 12--20 keV
energy band.  In Table \ref{table:log} we report the burst start time
detected with {\it INTEGRAL} from February 2003 to January 2005. The
start time for each burst was determined when the intensity was 10\% of the
peak above the persistent intensity level. Note
that six of the IBAS detected bursts were outside of the JEM-X field of view.  
Due to the lack of the low energy data we excluded these bursts
  from the analysis. However we report the burst start times and
  durations (see Table \ref{table:log}).    

\section{Results}
\label{sec:res}

\begin{figure}
\centerline{\epsfig{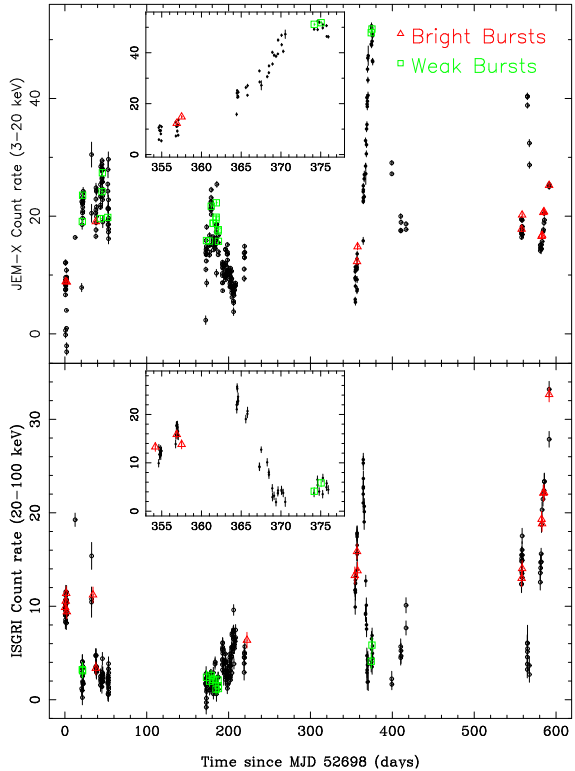}}
\caption
{JEM-X and IBIS/ISGRI light curves of \gx\ in the 3--20 keV and  20--100 keV
  energy bands. We indicate with red and green the net
  intensity (burst subtracted) emission where bursts were
  emitted. Each point corresponds to a pointing of $\sim$ 1800 s
  duration. The 21-day flare is shown on a smaller time scale in
  the insets.}    
\label{fig:isgri_lc}
\end{figure}

To study the light curves and spectra of \gx\ we
first deconvolved and analyzed separately the 356 pointings and
then combined them into a total mosaic image in the 20--40 keV energy
band. In the mosaic \gx\ is clearly detected, as well as the nearby
source the Rapid Burster (4U 1730-335).  
The source position offsets with respect to the catalog positions are
$0.\!\arcmin2$ for \gx\ and $0.\!\arcmin23$ for the Rapid
Burster. This is within the 90\% confidence level assuming the source
location error given by \citet{gros03}. The derived angular distance
between the two sources is $\sim31.\!\arcmin5$.  Due to the fact that
{\em INTEGRAL} is able to image the  
sky at high  angular resolution ($12\arcmin$ for ISGRI and $3\arcmin$ for 
JEM-X), we were able to clearly distinguish and isolate the 
high-energy fluxes from the two sources.  This allowed us to study 
 the X-ray emission of \gx\ during its entire observation without
 contamination \citep[see also for the Rapid Burster][]{falanga04}. 
In Fig. \ref{fig:isgri_lc} we show the JEM-X (top) 3--20 keV and
ISGRI (bottom) 20--100 keV light curves extracted from all the 356
images. Pointings during which bursts were detected are
marked in red (bright burst) and green (weak burst). The source is
very active and strong 
variations on a short time scale are not uncommon. Bursts were observed
during all intensity states. We also observed  an
isolated strong flare (around 53058 MJD), during which the
source count rate increases from $\sim$ 5 cts s$^{-1}$ to $\sim$ 50
cts s$^{-1}$ in the soft energy band, 3--20 keV, and shows a different
trend at higher energy, 20--100 keV. During this flare, lasting $\sim21$
days, the source undergoes a transition from a hard to a soft state (see
Sect. \ref{sec:spectrum}). Note that during this flare two weak bursts
happen at the peak of the 3--20 keV flux.

\begin{table*}[htb]
\caption{Log of {\it INTEGRAL} burst observations analyzed in this
  paper and spectral properties of the bursts. 
}
\begin{center}
\begin{tabular}{lllllllllll}
\hline 
Burst & IBIS/JEMX  & IBIS/JEM-X &  IBIS  & JEMX & & {\sc Burst}&
\multicolumn{2}{c} {\sc Parameters} & \\
Nr.  & Date & T$_{\rm start}$ (UTC) & $\Delta$T$_{\rm Burst}$
 & $\tau_{\rm exp}$ & $\dot m$ & E$_{b}$ & F$_{\rm
  peak}$& $\tau$ & $\gamma$ & $\alpha$\\
\hline
1$^{\dagger}$ & 2003 Feb 28 & 07:55:01 & 8.4(3) &--&--&--&--&--&--&--\\
2$^{\dagger}$ & 2003 Mar 01 & 00:04:45 & 8.2(3) & 6.8(5) & 1.7(2) & 8.3(3)& 9.6(3)& 8.6(6) & 0.012(4)&$\sim110$\\
3$^{\dagger}$ & 2003 Mar 01 & 16:05:27 & 9.4(3) &--&--&--&--&--&--&--\\
4$^{\dagger}$ & 2003 Mar 02 & 07:42:17 & 11.0(3)& 6.8(5) & 1.7(2) & 8.9(3)& 9.9(3)&9.0(4)&0.016(3)&$\sim100$\\
5             & 2003 Mar 21 & 06:09:48 & 5.7(3) & 4.5(5) & 3.2(2) & 3.5(3)& 3.9(3)&9.0(9)&0.07(7)&--\\
6             & 2003 Mar 21 & 15:46:38 & 3.5(3) & 3.6(5) & 4.3(2) & 3.2(2)& 7.9(2)&4.1(3)&0.051(2)&$\sim270$\\
7$^{\dagger}$ & 2003 Apr 03 & 08:40:16 & 4.2(3) &--&--&--&--&--&--&--\\
8$^{\dagger}$ & 2003 Apr 06 & 19:45:28 & 5.8(3) & 5.9(5) & 3.7(2) & 5.1(4)& 7.4(4)&6.9(6)&0.046(8)&--\\
9             & 2003 Apr 12 & 07:12:20 & $\star$& 3.2(5) & 3.5(2) & 2.6(2)& 4.4(2)&5.8(5)&0.073(6)&--\\
10            & 2003 Apr 14 & 09:05:15 & $\star$& 3.4(5) & 3.7(2) & 1.9(2)& 3.4(2)&5.5(7)&0.103(8)&--\\
11            & 2003 Apr 14 & 20:13:47 & $\star$& 5.0(5) & 4.5(3) & 2.7(3)& 3.8(2)&7.1(9)&0.111(8)&$\sim610$\\
12            & 2003 Apr 21 & 09:31:41 & $\star$& 5.4(5) & 3.7(2) & 2.1(2)& 3.6(2)&5.8(6)&0.062(7)&--\\
13            & 2003 Apr 19 & 22:57:35 & 3.5(3) & 3.4(5) & 2.7(2) & 3.1(2)& 5.1(2)&6.1(5)&0.049(4)&--\\
14            & 2003 Aug 23 & 16:14:00 & 4.2(3) & 5.0(5) & 2.7(2) & 3.5(2)& 6.1(2)&5.7(4)&0.042(4)&--\\
15            & 2003 Aug 25 & 18:24:08 & 6.0(3) & 4.1(5) & 3.9(2) & 3.6(2)& 6.6(2)&5.5(3)&0.056(3)&--\\
16            & 2003 Aug 28 & 01:23:58 & 7.6(3) & 3.6(5) & 3.2(5) & 3.9(5)& 5.5(5)&7.1(6)&0.055(5)&--\\
17            & 2003 Aug 28 & 06:01:29 & 4.6(3) & 2.6(5) & 3.3(5) & 3.1(5)& 5.3(5)&5.8(6)&0.057(6)&$\sim160$\\
18            & 2003 Aug 31 & 15:54:14 & 6.8(3) & 6.5(5) & 3.1(4) & 3.1(4)& 6.3(4)&5.0(7)&0.046(7)&--\\
19            & 2003 Aug 31 & 20:00:09 & $\star$& 4.0(3) & 3.5(3) & 1.9(4)& 3.9(4)&4.9(9)&0.08(1)&$\sim240$\\
20            & 2003 Aug 31 & 23:48:20 & $\star$& 5.6(3) & 4.6(2) & 3.1(3)& 3.7(3)&8.4(9)&0.12(1)&$\sim200$\\
21            & 2003 Sep 03 & 03:26:33 & 5.0(3) & 3.1(5) & 3.1(3) & 3.8(5)& 7.1(5)&5.3(8)&0.04(7)&--\\
22            & 2003 Sep 03 & 08:39:29 & 6.8(3) & 4.9(5) & 2.9(5) & 2.8(3)& 4.5(3)&6.2(7)&0.06(5)&$\sim170$\\
23            & 2003 Sep 03 & 13:07:21 & 2.7(3) & 3.9(5) & 2.8(4) & 2.1(4)& 4.4(4)&4.8(9)&0.06(1)&$\sim220$\\
24$^{\dagger}$& 2003 Oct 08 & 09:58:35 & 5.5(3) &--&--&--&--&--&--&--\\
25$^{\dagger}$& 2004 Feb 17 & 04:47:27 & 9.9(3) &--&--&--&--&--&--&--\\
26$^{\dagger}$& 2004 Feb 19 & 21:06:41 & 5.6(3) & 1.97(5) & 2.4(4)& 4.2(4)& 8.2(4)&5.1(5)&0.028(5)&--\\
27$^{\dagger}$& 2004 Feb 20 & 12:00:21 & 6.3(5) &--&--&--&--&--&--&--\\
28            & 2004 Mar 08 & 04:14:44 & 4.5(3) & 5.2(5) & 12.4(3)& 3.3(3)& 6.8(3)&4.8(6)&0.171(5)&--\\
29            & 2004 Mar 09 & 00:58:20 & 4.9(3) & 3.6(5) & 10.4(4)& 3.5(4)& 6.9(4)&5.0(7)&0.14(5)&$\sim690$\\
30            & 2004 Sep 07 & 21:30:45 & 5.0(3) & 3.1(5) & 3.8(3) & 4.2(3)& 7.0(3)&6.0(6)&0.056(5)&--\\  
31$^{\dagger}$& 2004 Sep 08 & 12:44:37 & 6.8(3) & 4.9(5) & 4.1(3) & 4.6(3)& 7.3(3)&6.3(6)&0.052(5)&$\sim450$\\
32$^{\dagger}$& 2004 Oct 02 & 05:59:14 & 2.7(3) & 3.9(5) & 4.1(3) & 4.1(3)& 6.9(3)&5.9(6)&0.054(6)&--\\
33$^{\dagger}$& 2004 Oct 02 & 21:51:07 & 6.1(3) & 5.5(5) & 3.7(4) & 3.6(4)& 7.7(4)&4.7(5)&0.074(5)&$\sim550$\\
34$^{\dagger}$& 2004 Oct 04 & 12:29:37 & 4.5(3) & 5.2(5) & 4.7(3) & 3.9(3)& 7.3(3)&5.3(5)&0.059(6)&--\\
35$^{\dagger}$& 2004 Oct 05 & 05:59:09 & 6.0(3) & 4.6(5) & 5.5(3) & 4.0(3)& 6.7(3)&5.9(5)&0.082(5)&$\sim780$\\
36$^{\dagger}$& 2004 Oct 11 & 12:59:09 & 4.9(3) & 6.8(5) & 1.7(3) & 3.5(3)& 7.0(3)&5.0(5)&0.091(6)&--\\
%13 &028600490010&2005 Feb 16 & 09:29:22&\multicolumn{2}{c}{Not public data}\\ 
%14 &028600580010&2005 Feb 16 & 13:54:59&\multicolumn{2}{c}{Not public data}\\ 
%15 &028600680010&2005 Feb 16 & 19:17:58&\multicolumn{2}{c}{Not public data}\\ 
%16 &029800840010&2005 Mar 24 & 21:14:26&\multicolumn{2}{c}{Not public data}\\ 
%17 &030400020010&2005 Apr 09 & 21:34:09&\multicolumn{2}{c}{Not public data}\\ 
\noalign{\smallskip} 
\hline 
\noalign{\smallskip} 
\end{tabular}
\end{center}
Notes : $^{\dagger}$ bright type I X-ray bursts. $\star$
  not a statistically significant detection in ISGRI. The burst exponential
  decay time, $\tau_{\rm exp}$, is in units of seconds in the 3--20
  keV energy band. The mass accretion
  rate per unit area of
  the persistent emission before the bursts, $\dot m$, are expressed in
  units of $10^{3}$ g cm$^{-2}$ s$^{-1}$, assuming a canonical NS radius of 10 km. 
The net burst fluence, $E_{\rm b}$, and the net peak flux, $F_{\rm peak}$,
  (both persistent emission subtracted) are
  in units of $10^{-7}$ erg cm$^{-2}$ and $10^{-8}$ erg cm$^{-2}$
  s$^{-1}$, respectively. The  bursts parameters $\tau=E_{\rm
  b}/F_{\rm peak}$ is in units of seconds, 
and $\gamma=F_{\rm pers}/F_{\rm peak}$ is also listed for each
  burst. The quantity $\alpha$ is defined as the
  ratio of the total energy emitted in the persistent flux to that
  emitted in bursts,
  $\alpha=F_{\rm pers}\Delta t / E_{\rm b}$, where $\Delta t$ is the
  time interval from the beginning of one burst to the next. We report
  only the most confident $\alpha$ values, i.e. without data gap
  between two burst.
\label{table:log}
\end{table*}

\subsection{Burst light curves}
\label{sec:lc_burst}

\begin{figure*}
\centerline{\epsfig{file=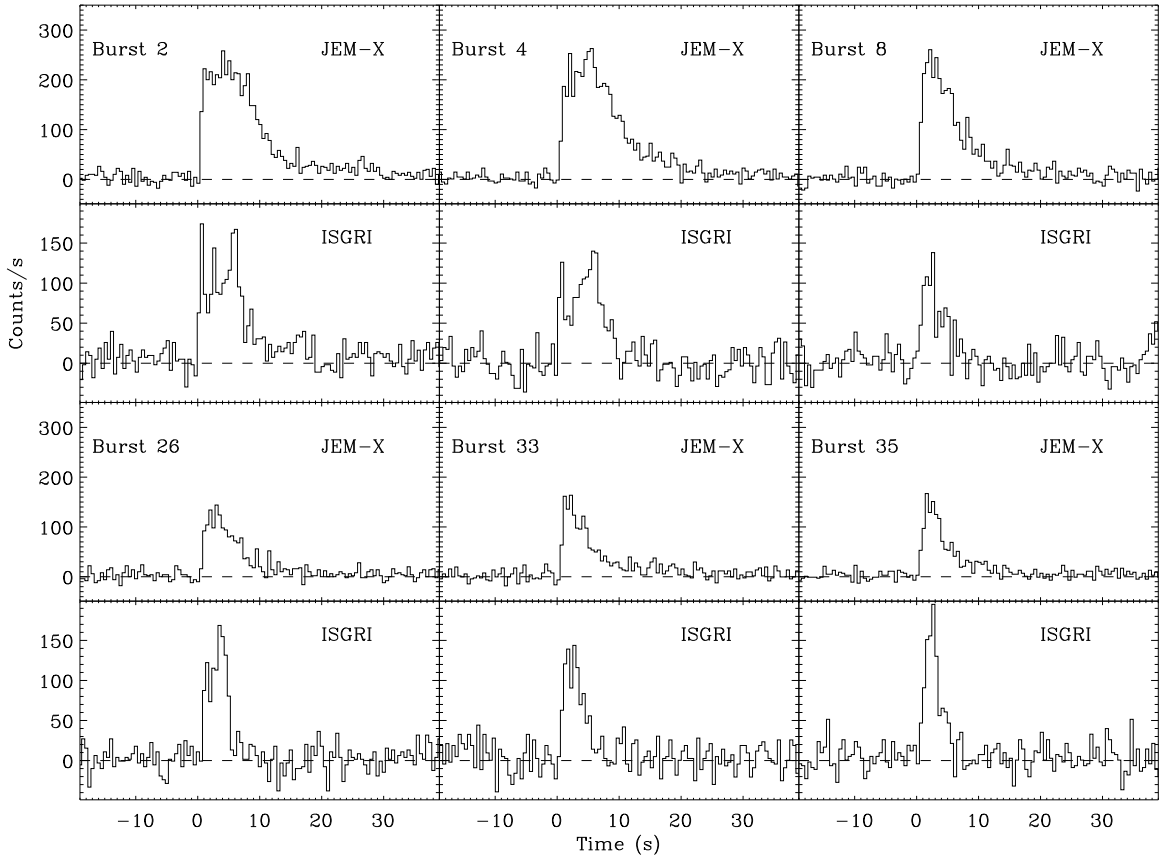,width=12.0cm,angle=90}}
\caption
{Some examples of bright type I X-ray bursts detected from \gx.
For each burst the JEM-X (3--20 keV, upper panels) and IBIS/ISGRI (18--40
keV, lower panels) net light curves are shown (background
subtracted). The time bin is 0.5 s for both IBIS/ISGRI and
JEM-X. Bursts 2 and 4 show evidence of photospheric radius expansion.
}
\label{fig:lcburst}
\end{figure*}

In total, we observed 36 bursts, which we label
numerically according to time. Their main properties are listed in
Table \ref{table:log}. Based on JEM-X/ISGRI light curves in the 3--6
keV, 6--12 keV, 12--20 keV and 
20--40 keV energy bands, rebinned at 0.5 s, we determined that the
bursts have rise times of $1\pm0.5$ s. The
decay time at 12--20 keV energy is significantly shorter than that in
the 3--6 keV range, indicating spectral softening during each burst
decay. This was also verified  by determining 
hardness ratios, using the three JEM-X energy bands. For JEM-X the flux
decays to quiescence with an mean exponential time-scale $F \propto
e^{-t/4.5^{s}}$ (see Table \ref{table:log}). At high energy, the bursts
were only significantly detected in the first $\Delta$T$_{\rm
  Burst}\sim$ 5--10 s (see Table \ref{table:log}).  
In Fig. \ref{fig:lcburst} we show six JEM-X/ISGRI representative burst
light curves in different energy bands.
For bursts 2 and 4, a double peak profile is evident at
high energy (lower panel) within the first 10 s, while 
during this time the intensity at lower energy (upper panel) remains constant. 
This can be interpreted as a consequence of a photospheric radius
expansion (PRE) episode during the  first part of the outburst \citep[see
  e.g.,][]{vh95}. When a burst undergoes a PRE episode,
due to radiation pressure, the luminosity, $L$, remains constant at
the Eddington value, $L_{\rm Edd}$, and, at high energy, a 
double peak profile can be observed. 
Such  behaviour is well known for \gx\, and was observed previously in
different bursts and recently reported with {\it RXTE} data
\citep{franco01,vss01,galloway03,galloway06}. 

We attempted to test the hypothesis of a PRE using the four different
JEM-X/ISGRI energy bands and we measured such a double peak behaviour
in the hardness ratio only for burst 2 and 4. 

\subsection{Hardness Intensity Diagram of the persistent emission}
\label{sec:HID}

\begin{figure}
\centerline{\epsfig{file=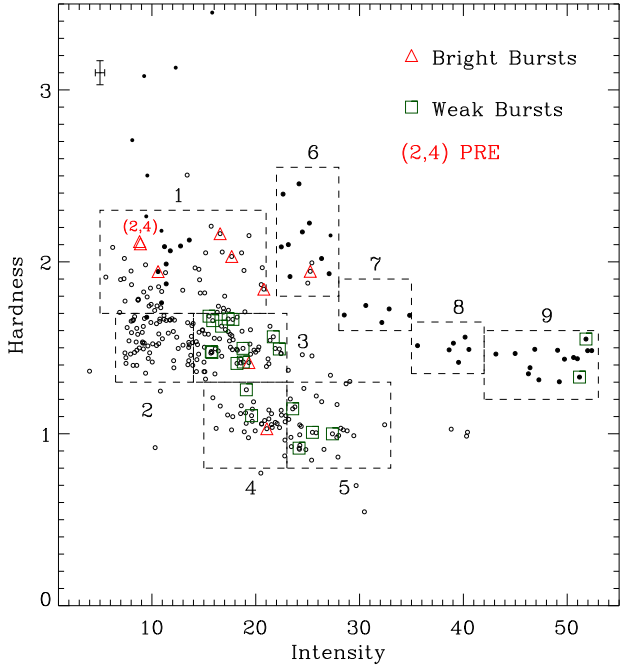,width=9.5cm}}
\caption
{HID of \gx. The hardness is the ratio of the count rates in the 6--20
 keV to 3--6 keV and the intensity is the 3--20 keV count rate. Each
 point (burst-subtracted) corresponds to $\sim 1800$ s of integration 
 time. A
 typical error bar is shown at the top left corner.
The boxes represent our selection of the data for the spectral
 analysis. The choice of the boxes (1--5 and 6--9) was made in order to
 study the spectral evolution along the atoll and to compare our result with
 previously studied \gx\ spectra. The
 filled dotted points are from the flare with $I_{\rm max}/I_{\rm min}
 \sim 10$ in the 3-20 keV energy band which lasted 21 days (see
 Fig. \ref{fig:isgri_lc}). The  bursts detected with JEM-X and ISGRI are
 also indicated. Bursts 2 and 4 show evidence of a photospheric radius
 expansion episode. 
}
\label{fig:HID}
\end{figure}

To generate a CCD \citep[see for this source e.g.,][]{hasinger89,salvo01,franco01} we used the background subtracted and burst
subtracted JEM-X light curve with a 500 s time resolution. We defined
the soft colour as the logarithm of the ratio of the count rate in the
energy range 3--4.8 keV to the count rate in the range 4.8--7.7 keV. 
Similarly, we defined the hard colour in the range 7.7--12.5 keV and
12.5--20 keV.   
In this case the CCD was inadequate as the colour errors were as
large as the colour variation. In order to reduce the errors on the
colours, we generated a HID based on the net count rates in the 3--6 keV and
6--20 keV obtained in the single individual pointings of $\sim$ 1800 s
each. In our case the source behaviour is much
better traced in the HID rather than in the CCD, because the
statistical uncertainties are reduced along one of the two axes.
The HID is shown in 
Fig. \ref{fig:HID}, with the positions superposed where the bursts were
detected. Note that the patterns observed in a CCD are often also
recognizable in the corresponding HID obtained by replacing soft colour by
intensity, as applied e.g. on the Z sources GX 17+2 \citep[][]{disalvo00,
farinelli05}, GX 5-1 \citep{munoa02} and on 
the atoll source Aql X-1 \citep{munoa02}. 
This results from the  intensity being dominated in the
lower bands by soft photons whose thermal component correlates with
temperature and therefore colour. The HID could indicate the banana
branch of the atoll, with inferred accretion rate increasing along the atoll
from left to right on the upper branch to the lower banana branch.   
To study the spectral variability of the source as a function of the
HID position we divided the HID into nine regions, as shown in
Fig. \ref{fig:HID}.  The intervals were chosen in order to have  
good statistics for the spectral analysis. We averaged the different
intensities with nearly the same hardness ratios and considering the nearness
colours these spectra should be very similar.  We also studied in more
detail the 21 day long flare (see Fig. \ref{fig:isgri_lc}), divided
among the boxes 6--9, represented by the filled 
dotted points in Fig. \ref{fig:HID}. The HID of the flare shows a
transition from a hard to a soft state during 21
days with an intensity ratio $I_{\rm  max}/I_{\rm min} \sim 10$ (see
Sect. \ref{sec:discussion}). This trend held also for the persistent
emission, which also moved from a hard to soft state (boxes
1--5). This overall trend can also be seen in the {\it RXTE}/ASM light curve
(Fig. \ref{fig:lc_asm}) for the period of this observation.

\subsection{Spectral Analysis}
\label{sec:spectrum}

The spectral analysis was done using XSPEC version 11.3 \citep{arnaud96},
combining the 20--200 keV ISGRI data with the simultaneous 5--20 keV
JEM-X data.  
A constant factor was included in the fit to take into account the
uncertainty in the cross-calibration of the instruments. The factor
was fixed at 1 for the ISGRI data. 
A systematic error of 2\% was applied to JEM-X/ISGRI spectra which
corresponds to the current uncertainty in the response matrix.
All uncertainties in the spectral parameters are given at a 90\%
confidence level for single parameters. For the source distance we use
4.5 kpc \citep{galloway03,galloway06}.   

\subsubsection{Persistent emission variability}
\label{sec:spec_pers}
We studied the spectra of \gx\ along the atoll pattern for
the different boxes reported in Fig. \ref{fig:HID}. 
A broad band 3--200 keV spectrum was obtained for each box using the
joint JEM-X/ISGRI data and removing the time intervals corresponding to
the bursts.

\begin{figure*}
\centerline{\epsfig{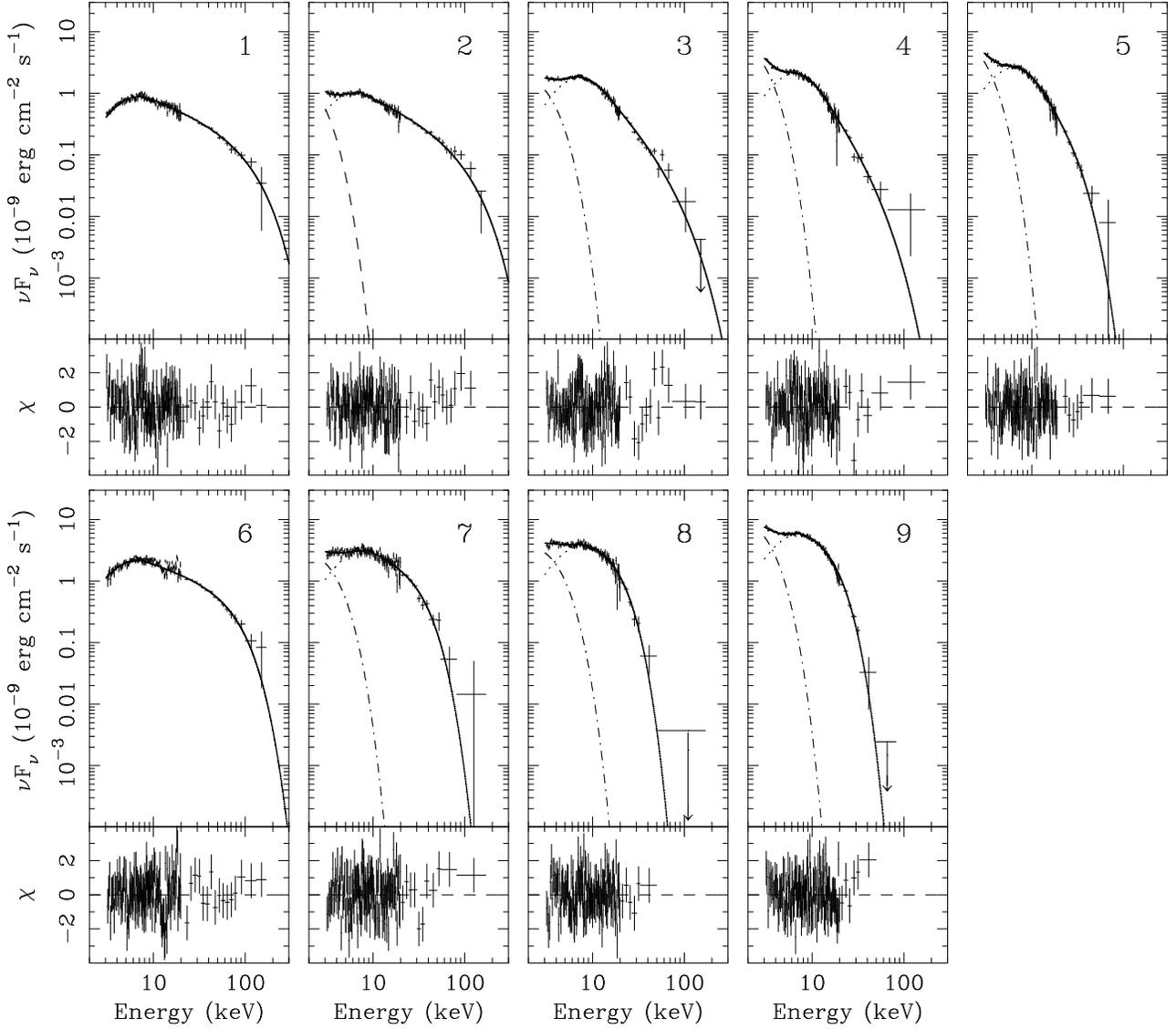}}
\caption
{{\em INTEGRAL}/JEM-X (3--20 keV) and ISGRI (20--200 keV) unfolded spectra of
      \gx\ along the ``banana'' track with the best fit \dbb\ and
      \comptt\ models. Residuals between the data and model are shown in the
      bottom panel in units of sigma. The spectra are labeled
      corresponding to each box number selected in the HID in
      Fig. \ref{fig:HID}. The best fit parameters are 
      shown in Fig. \ref{fig:spec_evol} and Table \ref{table:spec}.} 
\label{fig:continuo}
\end{figure*}

\begin{figure}
\centerline{\epsfig{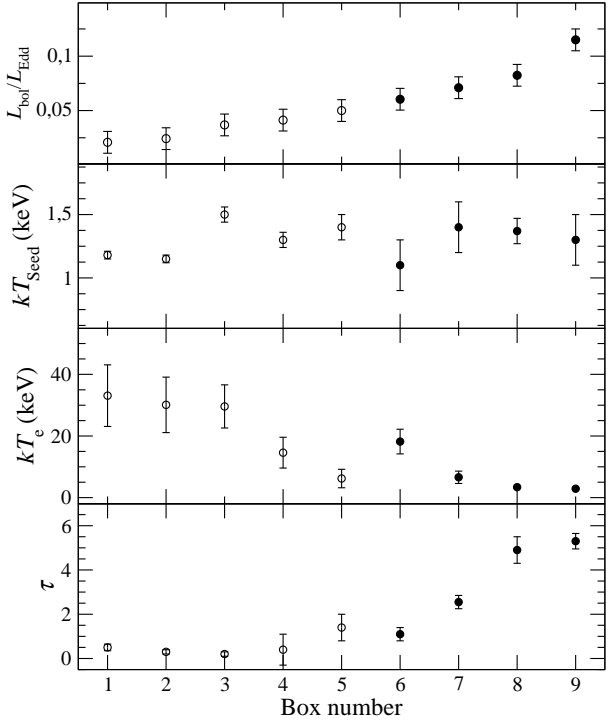}}
\caption
{Spectral parameters from Table \ref{table:spec} as a function of the
  position on the HID.}
\label{fig:spec_evol}
\end{figure}

Atoll source spectra are successfully fitted by a two-component model
consisting of a multicolour blackbody soft X-ray emission and a
Comptonized spectrum, for the hard X-ray emission. The emission from
the accretion shock 
on a neutron star is expected to be produced by thermal Comptonization 
of soft seed photons from the star, or from an optically thick
boundary layer between accretion disc and NS surface
\citep[e.g.,][]{zs69,alme73}. We model the shock emission by the
{\sc comptt} model \citep{titarchuk94} which is an  analytic model 
describing Comptonization of soft photons in a hot plasma. We
approximate the accretion shock geometry by a plane-parallel slab at
the neutron star surface. The main model parameters are the Thomson
optical depth across the slab, $\tau_{\rm T}$, the electron temperature,
$kT_{\rm e}$, and the seed photon temperature, $kT_{\rm seed}$. The soft
thermal emission, $kT_{\rm soft}$, believed to be produced by the disc,
is fitted by a multi-temperature disc model of \citep{mitsuda84}.  
We were not able to constrain the interstellar column density {\nh}
(as the JEM-X bandpass starts  
above 3 keV), so we have set {\nh} to the value 2.5 $\times 10^{22} {\rm
  cm}^{-2}$ observed by {\it ROSAT} and {\it BeppoSAX}.  
We add to the fit a Gaussian line around 6.4 keV with a fixed width at
$\sigma= 0.34$ keV as observed previously \citep[e.g.,][]{salvo00,piraino00}, 
but it was not statistically required. The multi-temperature disc model was
also not required in the fit during the low/hard state. Soft emission
(presumably from the accretion disc) during this state is difficult to
detect since the JEM-X bandpass begins at 3 keV. Compton reflection
between 10--30 keV was not required by the fit. The best fit
parameters for each box are reported in Table \ref{table:spec} and in
Fig. \ref{fig:spec_evol}. For all the fits, the normalization
  constants of the JEM-X response were within $1.08\pm0.09$. The nine
unabsorbed $\nu F_{\nu}$ spectra and the residuals  of the data to the
model are shown in Fig. \ref{fig:continuo}.  

\begin{table*}[htb]
\caption{Fit parameters of the persistent spectra in the energy range
  3--200 keV}
\begin{center}
\begin{tabular}{llllllllll}
\hline 
& Box 1 & Box 2 & Box 3  & Box 4 & Box 5 & Box 6 & Box 7 & Box 8 & Box 9\\
\hline
Parameter  & & &\multicolumn{4}{c}  \mbox{({\sc diskbb} + \comptt}) &  &\\
\hline
\noalign{\smallskip}
%$N_{\rm H} (10^{22} {\rm cm}^{-2})$    & 2.5 (f) & 2.5 (f) & 2.5 (f) &
%             2.5 (f) & 2.5 (f) & 2.5 (f) & 2.5 (f) & 2.5 (f) & 2.5 (f)\\ 
$kT_{\rm soft}$ (kev)&  --&0.54$^{+0.3}_{-0.4}$& 0.7$^{+0.3}_{-0.1}$&0.6$^{+0.5}_{-0.4}$  &  0.6$^{+0.3}_{-0.2}$ & --& 0.73$^{+0.23}_{-0.16}$ & 0.78$^{+0.12}_{-0.14}$& 0.65$^{+0.15}_{-0.16}$\\
$N_{\rm soft}^{1/2}$  &--& 40$^{+12}_{-12}$ & 20$^{+5}_{-4}$& 62$^{+15}_{-14}$ & 58$^{+11}_{-12}$& --& 30$^{+9}_{-10}$ & 27$^{+8}_{-8}$&69$^{+17}_{-18}$\\
%$E_{\rm Fe}$ (keV) & 7.3$^{+0.1}_{-0.4}$ & 7.07$^{+0.2}_{-0.4}$ &7.1$^{+0.15}_{-0.15}$& -- & 7.2$^{+0.3}_{-0.5}$& 7.3$^{+0.2}_{-0.3}$& -- & --& 6.8$^{+0.4}_{-0.5}$\\
%$\sigma_{\rm Fe}$ (keV) & 0.34 (f) & 0.34 (f) & 0.34 (f)& --& 0.34 (f) & 0.34 (f) & --&--& 0.34 (f)\\
$kT_{\rm seed}$ (keV)& 1.18$^{+0.03}_{-0.03}$
&1.15$^{+0.04}_{-0.03}$&1.5$^{+0.05}_{-0.07}$& 1.3$^{+0.06}_{-0.06}$ &
1.4$^{+0.08}_{-0.2}$ & 1.1$^{+0.3}_{-0.2}$  & 1.4$^{+0.22}_{-0.16}$ & 1.37$^{+0.1}_{-0.14}$& 1.3$^{+0.3}_{-0.13}$\\
$kT_{\rm e}$ (keV)& 34.9$^{+14}_{-6}$ & 32.1$^{+12}_{-7}$ &29.6$^{+4}_{-10}$&14.6$^{+33}_{-7}$ & 6.2$^{+4.7}_{-1.1}$ & 18.2$^{+6}_{-2}$& 6.6$^{+1.2}_{-0.75}$ & 3.4$^{+0.29}_{-0.23}$& 2.9$^{+0.12}_{-0.12}$\\ 
$\tau_{\rm T}$ & 0.5$^{+0.16}_{-0.2}$ & 0.3$^{+0.1}_{-0.1}$& 0.2$^{+0.2}_{-0.07}$&0.4$^{+1.2}_{-0.3}$ & 1.4$^{+0.7}_{-0.5}$ & 1.1$^{+0.2}_{-0.6}$& 2.55$^{+0.34}_{-0.4}$ & 4.9$^{+0.58}_{-0.56}$& 5.3$^{+0.4}_{-0.3}$\\ 
$\chi^{2}/{\rm dof}$  &  146/133 & 144/144& 150/139& 149/134 & 123/131 & 160/134& 160/136& 129/130& 126/133\\  
$L_{\rm 3-200keV}^{a}$ (erg s$^{-1}$) &3.8$\times10^{36}$ & 4.4$\times10^{36}$ & 6.7$\times10^{36}$ & 7.5$\times10^{36}$& 9.1$\times10^{36}$&
1.1$\times10^{37}$ & 1.3$\times10^{37}$& 1.5$\times10^{37}$ & 2.1$\times10^{37}$\\ 
\noalign{\smallskip}
\hline
\end{tabular}
\end{center}
$^{a}$ Assuming a distance of 4.5 kpc. The \dbb\ normalization is $N_{\rm
  soft} \propto R_{\rm in}^{1/2}$.  
\label{table:spec}
\end{table*}

\subsubsection{X-ray burst spectra}
\label{sec:spec_pers_burst}

To study the net bursts' spectra, we first extracted the
persistent emission spectrum excluding the outburst interval.  We
verified that during each burst pointing the count rate was stable,
then we used the same fit models as described in
Sect. \ref{sec:spec_pers} to fit the data. In Table \ref{table:log} we
report the mass accretion rate per unit area, $\dot m$, of the persistent
emission for each burst pointing. This is given by $\dot m = \dot
M/A_{\rm acc}$, where A$_{\rm acc}$ is the area covered by
  $4\pi R_{\rm NS}^{2}$. We assume a 10 km radius for the neutron star. The
persistent mass accretion rate, $\dot M$,  is 
calculated from $\dot M = L_{\rm pers}\;\eta^{-1}$ c$^{-2}$,
where $\eta \sim 0.2$ is the accretion efficiency for a NS. 
The bursts occurred at a persistent emission level $L_{\rm
  pers} \sim$ 2--8\% $L_{\rm Edd}$.

We then studied the net bursts' time averaged spectra using for each burst the
persistent emission spectra as background. The low statistics of
JEM-X/ISGRI do not allow us to study the spectral evolution
during the bursts. The spectra were extracted over the whole duration
of each burst as determined from the JEM-X light curves.  We also
included in the fit  the ISGRI spectra in the 18--40 keV energy bands. 
All obtained spectra were fitted by a photoelectrically absorbed
blackbody. The unabsorbed flux between 0.1--200 keV,
$F_{\rm bol,bb}$, was calculated according to $F_{\rm bol,bb} = 1.0763
\times 10^{-11} T^{-4}_{\rm bb}K^{2}_{\rm bb}$ erg cm$^{-2}$ s$^{-1}$,
where $T_{\rm bb}$ is the blackbody temperature in unit of
keV and $K_{\rm bb}(D,R_{\rm NS})$ is the
normalization of the blackbody component as returned by the fitting
program. The blackbody temperature of the time averaged burst
spectrum, kT$_{\rm bb}$, was between 1.8 keV and 2.4 keV. The burst
fluence, $E_{\rm b}$, is calculated by integrating the measured
$F_{\rm bol,bb}$ over the burst duration. The burst peak flux, $F_{\rm
  peak}$, was calculated by converting the peak count rate to flux
using the averaged spectrum of each burst. Taking into account the
typical blackbody temperature variation within the bursts
\citep{vss01}, the associated error on our conversion factor is of the
order of 12\%, which is within our statistical errors.  
The  best fit results and
calculated burst parameters are reported in Table \ref{table:log}.

\section{Discussion} 
\label{sec:discussion}

\subsection{Origin of X-ray emission} 
\label{sec:emission} 

We have fitted the 3--200 keV spectra of the atoll source \gx\ along its
track on the HID by a two-component model: a thermal Comptonization
model together with a soft component. The spectra and the parameters
are shown in Fig. \ref{fig:continuo} and Fig. \ref{fig:spec_evol}, 
respectively (see Table \ref{table:spec}). The change in the
spectral shape between the hard 
state (box/spectrum 1) and the soft state (box/spectrum 5) is clearly
visible, as also during the flare (box/spectra 6--9), probably caused
by a dramatic change in the accretion flow geometry. 
 During the spectral evolution, spectra 1--5, at low
intensity the    
optical depth was $\tau_{\rm T}\sim 0.5$ with a high plasma
temperature $kT_{\rm e} \sim 35$ keV (box 1), and at higher intensity
the optical depth was $\tau_{\rm T}\sim 1.4$ with a plasma
temperature as low as $kT_{\rm e} \sim 6$ keV (box 5). During this spectral
change the luminosity increases from  $\sim 0.02$ to 0.05 $L_{\rm
  Edd}$, where the seed photons temperature was almost constant at
$kT_{\rm seed} \sim 1.3$ keV. The same behaviour is observed also during
the flare event (box 6--9) where the  plasma temperature decreases
from $\sim 18$ to 3 keV and the optical depth increases from 1 to 5
and the spectral change the luminosity increases from  $\sim 0.06$ to
0.12 $L_{\rm Edd}$. We observe a typical atoll spectral change 
from a low or intermediate/hard state to a high/soft state, where the
accretion rate increases from the hard state to the soft state.
The soft component could be associated with the radiation from the
accretion disc.  

With our coverage of luminosity variation, $F_{\rm max}/F_{\rm min} \approx
2.5$, observed during $\sim7.4$ days integration time between 2003
February and 2004 October we are probably not observing the full atoll track,
even during the flare (see Fig. \ref{fig:HID}) . We observe the source
most likely from a intermediate/hard state, upper banana, to the soft
state, lower banana state, and probably not during the island
state. If the source was in the island state, we could 
also expect that more then two bursts with PRE are observed (see
Sect. \ref{sec:burst}).  
The hard component observed during the hard/intermediate state is
explained with Comptonization of seed photons from a NS surface by
hot electrons in an inner optically thin accretion flow (or outer
boundary layer). The optically thin Comptonization medium has
temperature of $\geq 20$ keV. The observed plasma
temperature and optical depth during the spectra 1--3 are
typically hard/intermediate state parameters, only during the flare we observe
typically lower banana state parameters. In the lower banana branch spectra the
Comptonized component is much softer, with temperature of $\sim 6$ keV
and optical depth of $\gg 1$ as observed during the spectra 5--9.

From our fits  it results that the hard spectra can be described by unsaturated
Comptonization of soft photons in the hot $kT_{\rm e} \sim 20-30$ keV
optically thin $\tau \sim 0.5-1$ plasma. However,
different models have been proposed to explain the different emission
region and emission mechanism during the different spectral states
and evolution \citep{mitsuda89,zs69,alme73,wsp88}.  

If the soft photons are emitted from a disc, then the inner
disc radius can be estimated from the {\sc diskbb} model as  
\begin{equation}
R_{\rm in} \approx 0.76 \,N_{\rm
  soft}^{1/2}\,\frac{2.7}{\eta}\,\frac{4.5}{D}\biggl(\frac{f_{\rm
  col}}{1.8}\biggr)^{2} \biggl(\frac{0.5}{\cos\,i}\biggr)^{1/2} \mbox{km},
\end{equation}
where D is the source distance, $i$ is the inclination angle of the
system, $f_{\rm col}$ is the ratio of the colour to effective
temperature \citep{st95} and $\eta$ is the correction factor for the
inner torque-free boundary condition \citep[][$\eta$ = 2.7 for $R_{\rm
 in}\sim R_{\rm NS}$ and less for higher $R_{\rm in}$]{getal99}. With
$D=4.5$ kpc, $f_{\rm col}=1.8$ and using a mean inclination angle of
$i=60^{\circ}$ we find $R_{\rm in} = 15\pm3 \div 52\pm13$ km.    
For \gx\ the observed QPO are related on their position in the CCD/HID 
\citep[][]{mendez99,salvo01,migliari03}. Among the most frequently
used QPO models, the upper kHz QPO is related to the Kepler frequency,
i.e. $\nu_{\rm QPO} = \nu_{\rm Kepler}\approx$ 1184 Hz ($R_{\rm in}$/15
km)$^{-3/2}\,M^{1/2}_{\odot,1.4}$, at the innermost stable orbit of
the accretion disc \citep[for a review see][and reference therein]{vdk04}.  
Using our measured inner disc radius, $R_{\rm in}$, the upper kHz QPO
has to be in the frequency range from $1160\pm300$ Hz to $213\pm53$
Hz. These values are consistent with the upper kHz QPO values of \gx\
\citep{salvo01}. However, using this relation we are not able to
 exactly locate the source position in the CCD/HID.

We have measured the plasma temperature, optical depth, and flux of
\gx\ over a range of positions on the HID for the first time. We compare with
previously reported time averaged spectra
\citep{gh81,basinska84,w86,claret94,schulz99,salvo00,piraino00,narita01,dai06}.
Accounting for the  fact that these authors have reported time
averaged spectra sometimes over possibly different 
spectral states, and sometimes only in energy ranges less than 20 keV,
and insofar as the models used to fit the data are consistent, we
found that all the previous reported spectral parameters are
consistent with our measurements. 
For example, \citet{claret94}, using {\it SIGMA} (30-200 keV) data, measured
 plasma temperatures KT$_{\rm e}$ of $\sim 5$ and $\sim30$ keV in two different
source states. These correspond to our reported soft and hard
states. Also using a broad band spectrum, BeppoSAX observed the source
with kT$_{\rm e}\sim 7$ keV and $\tau \sim 5$, also consistent with
an intermediate soft state. We
would like to point out, in order to determine the source states with
more confidence, how important it is for \gx\ to have: $(i)$ a 
high energy spectrum above 20 keV, and $(ii)$ not to average the spectra  over
a long time period. $(i)$ using the JEM-X and ISGRI broad band spectrum (box
1) from 3--200 keV, in the intermediate/hard state,  the thermal Comptonization
plasma temperature was $\sim$ 35 keV, considerably higher than the
value ($\sim10$ keV) derived from the low energy 3--20 keV JEM-X data
alone. $(ii)$ a flux variation of a factor 1.12 will already change
the spectral parameters and a factor 2 will drastically change the
source state (see Table \ref{table:spec}). 
Note also, e.g., that during the flare we have 
$I_{\rm max}/I_{\rm min} \sim 10$  in the 3--20 keV band which
corresponds to a flux variation of $F_{\rm max}/F_{\rm min} \sim 2$ in
the 3--200 keV band. This shows again for \gx\ how important  the
energy contribution is above 20 keV.      

\subsection{Type I X-ray bursts}
\label{sec:burst}

\begin{figure}[htb]
\centerline{\epsfig{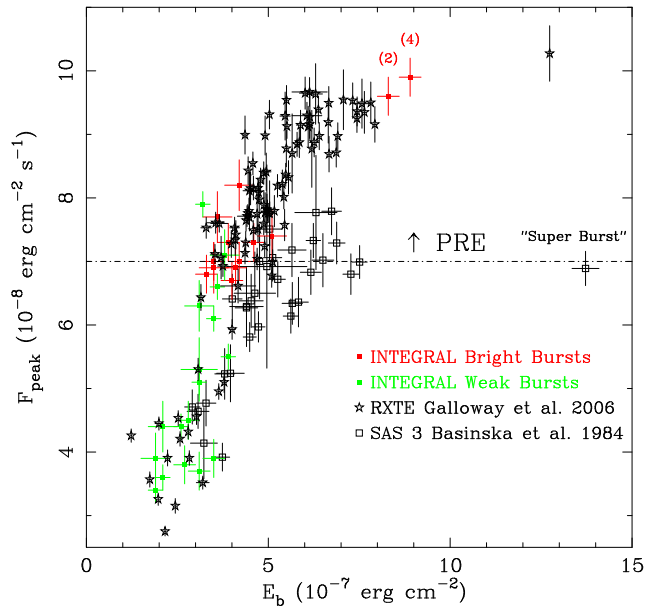}}
\caption
{Relation between peak flux, $F_{\rm peak}$, and burst fluence, $E_{\rm
  b}$. The dashed line indicates the lower limit of the observed
  photospheric radius expansion events. The two {\it INTEGRAL}
  bursts exhibiting PRE are labeled with the number 2 and 4 as reported in
  Table \ref{table:log}.}
\label{fig:spec_fb}
\end{figure}

More than 200 bursts have been observed in \gx\ since its discovery 
\citep{hoffman76,w86,basinska84,foster86,day90,franco01,vss01,galloway03,galloway06}.    
Compared to the previously observed bursts, ours appear as ordinary
bursts observed for this source. All the detected bursts have a rise
time about 1 s with a e-fold decay time from 3--7 s in the 3--20 keV
energy band. Most of the bursts are observed up to
30--40 keV during the first 5--10 s, and the energetic ones
are observed at the lowest inferred mass accretion rate (see
Fig. \ref{fig:HID}). Such behaviour was observed e.g., also for GX
3+1, where a super burst and a unusually long burst occurred at
the lowest mass accretion rate \citep{kvdk00,Chenevez06}. 

The averaged spectrum during the burst measured a
blackbody temperatures from 1.8 to 2.4 keV with a bolometric flux from
$1.7\times10^{-8}$ to $4.5\times10^{-8}$ erg cm$^{-2}$ s$^{-1}$,
respectively.  This corresponds to an effective emission radius 
(i.e., $R_{\rm eff}=d_{\rm 10kpc}\,F_{bol}^{1/2}\,T_{\rm bb}^{-2}\,\sigma^{-1/2}\,(z+1)^{-1/2}$ where $z+1=1.31$ is the gravitational
redshift correction and $\sigma$ is the Stefan-Boltzman constant), $R_{\rm
  eff}$, from $\sim4.5$ d$_{\rm 4.5 kpc}\div3.8$ d$_{\rm 4.5 kpc}$ km,
assuming a NS radius of 10 km, a NS mass of 1.4 M$_{\odot}$ and isotropic
emission. This emission radius could be attributed to an extended part
of the NS surface. 

\begin{figure}
\centerline{\epsfig{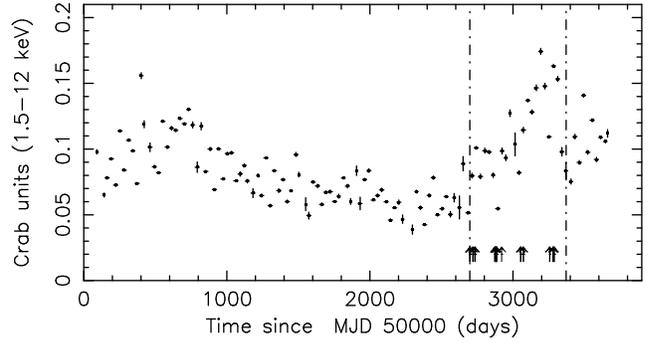}}
\caption{30-day averaged {\it RXTE}/ASM light curve in the 1.5--12
  keV energy band  showing the long term variability of \gx\ from 1996
  to 2005. The vertical dashed lines indicate the time interval of
  the {\it INTEGRAL} observations. The ASM count rates has been
 converted into flux using 1 Crab Unit for $75$ cts/s. The arrows
  indicate the time of the type I bursts.}
\label{fig:lc_asm}
\end{figure}

In Fig. \ref{fig:spec_fb} we show the bursts' peak
flux, $F_{\rm peak}$, plotted as a function of the integrated burst
flux, $E_{\rm b}$, including the 32 and 104 bursts observed by {\it SAS
  3} and {\it RXTE} \citep{basinska84,galloway06}. Note that a subset
of the {\it RXTE} bursts were reported and discussed in \citep{galloway03}.  
Most of the bursts with a higher peak flux as
$\sim7\times10^{-8}$ erg cm$^{-2}$ s$^{-1}$ show double-peaked
behaviour, these energy-dependent double-peaked profile are also the result of 
PRE. All the bursts below the dashed line are bursts without PRE. The
{\it INTEGRAL} 
bursts, excluding bursts with PRE,  occurred at a
persistent emission flux minimum and maximum level of
$\sim2.2\times10^{-9}$ and $\sim8.8\times10^{-9}$  erg cm$^{-2}$
s$^{-1}$ ($\dot m \sim 2.4\times10^{3} \div 9.4\times10^{3}$ g cm$^{-2}$
s$^{-1}$), respectively (3--20 keV). The values without PRE observed
by {\it SAS 3} \citep{basinska84} are within the {\it INTEGRAL} observation.  
The averaged value of the
persistent emission during the {\it SAS 3}  observation was
$\sim3.25\times10^{-9}$  erg cm$^{-2}$ s$^{-1}$ (1-20 keV) and varied
by a factor $\sim 1.25$. The {\it INTEGRAL} bursts with PRE, burst 2
and 4, occurred at the lowest mass accretion rate of
$\sim1.7\times10^{3}$ g cm$^{-2}$ s$^{-1}$. The bursts total fluence,
and the bursts peak flux are anti-correlated with the mass accretion rate.
From  a 30 days averaged {\it RXTE}/ASM light curve (1.5--12
 keV) \citep{l96} the persistent flux of \gx\ shows a variation of
 about a factor three on a $\sim7.7$ yr time scale and {\em
   INTEGRAL} observed \gx\ during a bright persistent emission phase (see
 Fig. \ref{fig:lc_asm}). Indeed the only two bursts with PRE where
 observed at the the lowest mass accretion rate at the start of the {\it
 INTEGRAL} observation, all other bursts without PRE are observed at higher
 mass accretion rate. 

The lowest inferred mass accretion rate at the $\sim7\times10^{-8}$
erg cm$^{-2}$ s$^{-1}$ line (see Fig. \ref{fig:spec_fb}) is $\dot
m\approx2.4\times10^{3}$ g cm$^{-2}$ s$^{-1}$.  
From the observed burst properties (see Table \ref{table:log}) and
inferred mass accretion rates, the present theory predicts that these
bursts are pure helium burning \citep[e.g.][]{s04}. For helium
flashes, the fuel burns rapidly, since there are no slow weak
interactions, and the local Eddington limit is often exceeded. These
conditions lead to PRE bursts with a duration, set mostly by the time
it takes the heat to escape, of the order of 5--10 s, as
observed. The {\it INTEGRAL}-observed bursts without PRE with a
mass accretion rate of $\gtrsim 2.4\times10^{3}$ g cm$^{-2}$ s$^{-1}$
could also arise from helium burning. In the framework of the
thermonuclear-flash models \citep[e.g.,][]{lewin95} the burst
duration, $\tau=E_{\rm b}/F_{\rm peak} < 10$ s, and the ratio of
the observed persistent flux to the net peak flux, $\gamma=F_{\rm pers}/F_{\rm
 peak}$, indicate a hydrogen-poor burst, i.e., pure helium bursts.  We
found a good agreement with the predicted amount of liberated helium
fuel during the bursts with the observed 
accreted mass between two bursts (calculated only for the bursts with
known $\alpha$ parameter).  The amount of fuel liberated in the
thermonuclear burning is calculated as L$_{b}/\epsilon_{\rm He}$, where
L$_{b}$ is the observed burst total fluence and $\epsilon_{\rm
He}\approx 1.7$ Mev/nucleon $\approx 1.6\times10^{18}$ erg g$^{-1}$ is the
He energy release. For instance, for burst 2, $\Delta M= \dot M \Delta t
\sim 1.2\times 10^{21}$ g is in agreement with the predicted amount of
liberated helium fuel during this burst L$_{b}/\epsilon_{\rm He}
\sim1.3\times10^{21}$ g. We observed pure helium bursts during the
%\approx2.01\times10^{39}/1.6\times10^{18}\sim1.3\times10^{21}$ g.
different spectral states. \gx\ shows the same burst properties as the
ultracompact 10 minute binary 4U 1820-30 \citep{cumming04}.  
In Fig. \ref{fig:spec_fb} we also show the so-called
``super burst'' by \citet{basinska84}. Super bursts, as currently
understood and observed, have a duration of several hours, a total
fluence of $\sim 10^{42}$ erg and are believed to be due to unstable
burning of a thick carbon layer \citep[e.g.][]{s04}. The observed
``super burst'' by \citet{basinska84} was most likely also a helium
burst with PRE triggered after a longer burst recurrence time.

\section{Summary} 
\label{sec:summary} 

We analyzed the spectral and bursting behaviour from the simultaneous  {\it
  INTEGRAL} IBIS/ISGRI and JEM-X observation of \gx, which was
spatially well distinguished from the neighbouring source, the Rapid Burster.
The broad band spectra from 3--200 keV is well described by a thermal
  Comptonization model with seed photons from the neutron star surface
  scattered in a shock-heated accretion column above the hot spot region,
or from an optically thick boundary layer between accretion disc and
  NS surface. The thermal blackbody soft emission could arise
from the accretion disc. The spectral evolution observed from an
  intermediate/hard state to the soft state are entirely typical of
  atoll sources. With increasing mass accretion rate the source
  spectra become brighter and softer.  We observed a decreasing electron
  temperature from 35--3 keV an increase of the optical depth
  from 0.5--5, during the source luminosity increase from 2-12\% of
  Eddington, i.e. from the intermediate/hard state to the soft state.  

The 36 type I bursts show an anti-correlation between burst peak flux
or fluence with the mass accretion rate. At low mass accretion rate,
intermediate/hard state, most of the bursts show PRE with a peak flux
exceeding the Eddington luminosity, at higher mass accretion rate, soft
state, the bursts were normal type I bursts. 
We observed normal X-ray bursts for this source involving pure helium
either during PRE in the intermediate/hard state, or during the soft
state.

\acknowledgements
MF and DG acknowledge the French Space Agency (CNES)  for financial
support. MF is grateful to D. Galloway for providing us with the {\it
  RXTE} data used in Figure 6 and for valuable discussions. 
The data reported in this paper are based on observations with
{\it INTEGRAL}, an ESA project with instruments and 
the science data center funded by ESA member states (especially the PI
countries: Denmark, France, Germany, Italy, Switzerland, Spain), Czech
Republic and Poland, and with the participation of Russia and the USA.
ISGRI has been realized and maintained in flight by CEA-Saclay/DAPNIA with the
support of CNES.

\end{document}